# The Rainbow Skip Graph:
# A Fault-Tolerant Constant-Degree P2P Relay Structure

Michael T. Goodrich[*]    Michael J. Nelson[*]    Jonathan Z. Sun[†]

**Abstract**

We present a distributed data structure, which we call the *rainbow skip graph*. To our knowledge, this is the first peer-to-peer data structure that simultaneously achieves high fault-tolerance, constant-sized nodes, and fast update and query times for ordered data. It is a non-trivial adaptation of the SkipNet/skip-graph structures of Harvey *et al.* and Aspnes and Shah, so as to provide fault-tolerance as these structures do, but to do so using constant-sized nodes, as in the family tree structure of Zatloukal and Harvey. It supports successor queries on a set of $n$ items using $O(\log n)$ messages with high probability, an improvement over the expected $O(\log n)$ messages of the family tree. Our structure achieves these results by using the following new constructs:

- *Rainbow connections*: parallel sets of pointers between related components of nodes, so as to achieve good connectivity between "adjacent" components, using constant-sized nodes.

- *Hydra components*: highly-connected, highly fault-tolerant components of constant-sized nodes, which will contain relatively large connected subcomponents even under the failure of a constant fraction of the nodes in the component.

We further augment the hydra components in the rainbow skip graph by using erasure-resilient codes to ensure that any large subcomponent of nodes in a hydra component is sufficient to reconstruct all the data stored in that component. By carefully maintaining the size of related components and hydra components to be $O(\log n)$, we are able to achieve fast times for updates and queries in the rainbow skip graph. In addition, we show how to make the communication complexity for updates and queries be worst case, at the expense of more conceptual complexity and a slight degradation in the node congestion of the data structure.

**Categories and Subject Descriptors:** E.2 Data Storage Representations

**General Terms:** Algorithms, Design.

**Keywords:** Distributed data structures, peer-to-peer networks, skip lists, skip graphs, family trees, erasure codes.

# 1 Introduction

Distributed peer-to-peer networks present a decentralized, distributed method of storing large data sets. Information is stored at the hosts in such a network and queries are performed by sending messages between hosts (sometimes iteratively with the query issuer), so as to ultimately identify the host(s) that store(s) the requested information. For the sake of efficiency, we desire that the assignment and indexing of data at the nodes of such a network be done to facilitate the following outcomes:

- *Small nodes*: Each node in the structure should be small. Ideally, each node should have constant size, including all of its pointers (which are pairs $(x, a)$, where $x$ is a host node and $a$ is an address on that node). This property allows for efficient space usage, even when many virtual nodes are aggregated into single physical hosts. (We make the simplifying assumption in this paper that there is a one-to-one correspondence between hosts and nodes, since a blocking strategy such as that done in the skip-webs framework of Arge *et al.* [2], can be used to assign virtual nodes to physical hosts.)

- *Fault tolerance*: The structure should adjust to the failure of some nodes, repairing the structure at small cost in such cases. Ideally, we should be able to recover the indexing data from failed nodes, so as to be able to answer queries with confidence.

- *Fast queries and updates*: The structure should support fast queries and insertions/deletions, in terms of the number of rounds of communication and number of messages that must be exchanged in order to complete requested operations. (We are not counting the internal computation time at hosts or the query/update issuer, as we expect that, in typical scenarios, message delays will be the efficiency bottleneck.)

- *Support for ordered data*: The structure should support queries that are based on an ordering of the data, such as nearest-neighbor searches and range queries. This feature allows for a richer set of queries than a simple dictionary that can only answer membership queries, including those arising in DNA databases, location-based services, and prefix searches for file names or data titles.

[*]Department of Computer Science, Bren School of Information and Computer Sciences, University of California, Irvine, CA 92697-3435. {goodrich,mjnelson}(at)ics.uci.edu.
[†]School of Computing, The University of Southern Mississippi, Hattiesburg, MS 39406-5106. jonathan.sun@usm.edu

To help quantify the above desired features, we use the following parameters, with respect to a distributed data structure storing a set $\mathcal{S}$ of $n$ items:

- $M$: the memory size of a host, which is the number of data items (keys), data structure nodes, pointers, and host IDs that any host can store.

- $Q(n)$: the *query cost*—the number of messages needed to process a query on $\mathcal{S}$.

- $U(n)$: the *update cost*—the number of messages needed to insert a new item in the set $\mathcal{S}$ or remove an item from the set $\mathcal{S}$.

- $C(n)$: the *congestion* per host—the maximum (taken over all nodes) of the expected fraction of $n$ random queries that visit any given node in the structure (so the congestion of a single distributed binary search tree is $\Theta(1)$ and the congestion of $n$ complete copies of the data set is $\Theta(1/n)$).

We assume that each host has a reference to the place where any search from that host should begin, i.e., a *starting node* for that host.

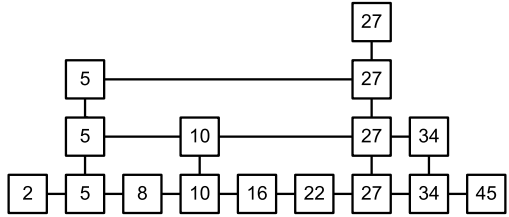

Figure 1: A skip list. Each element exists in the bottom-level list, and each node on one level is copied to the next higher level with probability 1/2. A search can start at any node and proceed up to the top level (moving left or right if a node is not copied higher), and then down to the bottom level. In the downward phase, we search for the query key on a given level and then move down to the next level, and continue searching until we reach the desired node on the bottom. The expected query time is $O(\log n)$ and the expected space is $O(n)$.

## 1.1 Previous Related Work

There is a significant and growing literature on distributed peer-to-peer data structures. For example, there is a considerable amount of work on variants of Distributed Hash Tables (DHTs), including Chord [9, 20], Koorde [13], Pastry [18], Scribe [19], Symphony [14], and Tapestry [22], to name just a few. Although they have excellent congestion properties, these structures do not allow for non-trivial queries on ordered data, such as nearest-neighbor searching, string prefix searching, or range queries. Aspnes and Shah [4] present a distributed data structure, called *skip graphs*, for searching ordered data in a peer-to-peer network, based on the randomized skip-list data structure [17]. (See Figure 1.) Harvey *et al.* [12] independently present a similar structure, which they call SkipNet. These structures achieve $O(\log n/n)$ congestion, expected $O(\log n)$ query time, and expected $O(\log n)$ update times, using $n$ hosts, each of size $O(\log n)$. Harvey and Munro [11] present a deterministic version of SkipNet, showing how to achieve worst-case $O(\log n)$ query times, albeit with increased update costs, which are $O(\log^2 n)$, and higher congestion, which is $O(\log n/n^{0.68})$. Zatloukal and Harvey [21] show how to modify SkipNet to construct a structure they call family trees, achieving $O(\log n)$ expected time for search and update, while restricting $M$ to be $O(1)$, which is optimal. Manku, Naor, and Wieder [15] show how to improve the expected query cost for searching skip graphs and SkipNet to $O(\log n/\log \log n)$ by having hosts store the pointers from their neighbors to their neighbor's neighbors (i.e., neighbors-of-neighbors (NoN) tables); see also Naor and Wieder [16]. Unfortunately, this improvement requires that the memory size and expected update time grow to be $O(\log^2 n)$, with a similar degradation in congestion, to $O(\log^2 n/n)$. Focusing instead on fault tolerance, Awerbuch and Scheideler [5] show how to combine a skip graph/SkipNet data structure with a DHT to achieve improved general fault tolerance for such structures, but at an expense of a logarithmic factor slow-down for queries and updates. Aspnes *et al.* [3] show how to trade-off the space complexity of the skip graph structure with its congestion, by bucketing intervals of keys on the "bottom level" of their structure. Their method reduces the overall space usage to $O(n)$, but increases the congestion to $O(\log^2 n/n)$ and still requires that $M$ be $O(\log n)$. Arge *et al.* [2] present a framework, called skip-webs, which generalizes the skip graph data structure to higher dimensions and achieves $O(\log n/\log \log n)$ query times, albeit with $M$ being $O(\log n)$ rather than constant-sized.

| Method | $M$ | $Q(n)$ | $U(n)$ | $C(n)$ |
|---|---|---|---|---|
| skip graphs/SkipNet [4, 12] | $O(\log n)$ | $O(\log n)$ w.h.p. | $O(\log n)$ w.h.p. | $O(\log n/n)$ |
| NoN skip-graphs [15, 16] | $O(\log^2 n)$ | $\tilde{O}(\log n/\log\log n)$ | $\tilde{O}(\log^2 n)$ | $O(\log^2 n/n)$ |
| family trees [21] | $O(1)$ | $\tilde{O}(\log n)$ | $\tilde{O}(\log n)$ | $O(\log n/n)$ |
| deterministic SkipNet [11] | $O(\log n)$ | $O(\log n)$ | $O(\log^2 n)$ | $O(n^{0.32}/n)$ |
| bucket skip graphs [3] | $O(\log n)$ | $\tilde{O}(\log n)$ | $\tilde{O}(\log n)$ | $O(\log^2 n/n)$ |
| skip-webs [2] | $O(\log n)$ | $\tilde{O}(\log n/\log\log n)$ | $\tilde{O}(\log n/\log\log n)$ | $O(\log n/n)$ |
| **rainbow skip graphs** | $O(1)$ | $O(\log n)$ w.h.p. | $O(\log n)$ amort. w.h.p. | $O(\log n/n)$ |
| **strong rainbow skip graphs** | $O(1)$ | $O(\log n)$ | $O(\log n)$ amort. | $O(n^\epsilon/n)$ |

Table 1: Comparison of rainbow skip graphs with related structures. We use $\tilde{O}(*)$ to denote an expected bound.

Thus, the family tree [21] is the only peer-to-peer structure we are familiar with that achieves efficient update and query times for ordered data while bounding $M$ to be $O(1)$ and maintaining a good congestion, which is $O(\log n/n)$. Unfortunately, Zatloukal and Harvey do not present any fault-tolerance properties of the family tree, and it seems difficult to do so.

## 1.2 Our Results

In this paper, we present *rainbow skip graphs*, which are an adaptation of the skip-graph of Aspnes and Shah [4] designed to reduce the size of each node to be $O(1)$ while nevertheless keeping the congestion at $O(\log n/n)$ and providing for improved fault tolerance. Successor queries use $O(\log n)$ messages with high probability, an improvement over the expected $O(\log n)$ messages of the family tree. The update and congestion complexities of our structure are also optimal (amortized in the update case), to within constant factors, under the restriction that nodes are of constant size. In addition, we present a strong version of rainbow skip graphs, which achieve good worst-case bounds for queries and updates (amortized in the update case), albeit at a slight decrease in congestion, which is nevertheless not as much as the decrease for deterministic SkipNet [11]. In Table 1, we highlight how our methods compare with previous related solutions.

Our improvements are based on the following:

- *Rainbow connections*: collections of parallel links between related components in the data structure. These connections allow for a high degree of connectivity between related components without the need to use more than a constant amount of memory per node.

- *Hydra components*: components of related nodes organized so that deleting even a constant fraction of the nodes in the component leaves a relatively large connected subcomponent.

We use the rainbow connections with the hydra components and erasure codes so that we can fully recover from significant sets of node deletions, even to recover all the lost data. We present a periodic failure recovery mechanism that can, with high probability, restore the correct structure even if each node fails independently with constant probability less than one. If $k$ nodes have failed, the repair mechanism uses $O(\min(n, k\log n))$ messages over $O(\log^2 n)$ rounds of message passing.

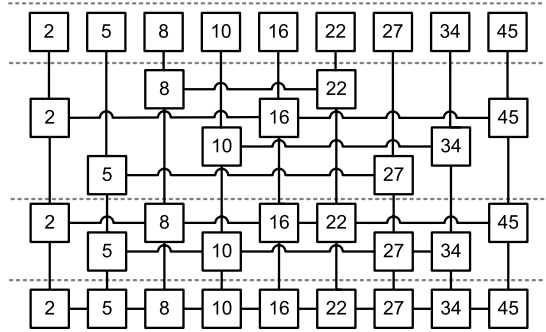

Figure 2: A skip graph. The dashed lines show the separations between the different levels.

## 2   Non-Redundant Rainbow Skip Graphs

Skip graphs [4, 12] can be viewed as a distributed extension of skip lists [17]. Both skip lists and skip graphs consist of a set of increasingly sparse doubly-linked lists ordered by levels starting at level 0, where membership of a particular node $x$ in a list at level $i$ is determined by the first $i$ bits of an infinite sequence of random bits associated with $x$, referred to as the *membership vector of $x$*, and denoted by $m(x)$. We further denote the first $i$ bits of $m(x)$ by $m(x)|i$. In the case of skip lists, level $i$ has only one list, for each $i$, which contains all elements $x$ s.t. $m(x)|i = 1^i$, i.e., all elements whose first $i$ coin flips all came up heads. As this leads to a bottleneck at the single node present in the uppermost list, skip *graphs* have $2^i$ lists at level $i$, which we will index from 0 to $2^i - 1$. Node $x$ belongs to the $j$th list of level $i$ if and only if $m(x)|i$ corresponds to the binary representation of $j$. Hence, each node is present in one list of every level until it eventually becomes the only member of a singleton list. (See Figure 2.)

It is useful to observe that the set of all lists to which a particular node $x$ belongs meets the definition of a skip list, with membership in level $i$ determined by comparison to $m(x)|i$ rather than to $1^i$. With this observation, the algorithms and time analysis[*] for searching, insertion, and well-behaved deletion in a skip graph all follow directly from the corresponding algorithms and analysis of skip lists. Nodes in a skip graph have out-degree proportional to the height of their corresponding skip list, which has been shown to be $\Theta(\log n)$ with high probability; thus, the storage requirement of each node includes $\Theta(\log n)$ address pointers to other nodes.

In the following subsection, we present a scheme that results in a new overlay structure, which we call a *non-redundant rainbow skip graph*. This structure has the property that each node has constant out-degree, and hence need store only a constant number of pointers to other nodes, matching the best-known results of the family tree [21]. Moreover, as we show, the non-redundant rainbow skip graph has $O(\log n/n)$ congestion. In subsequent sections, we show how to augment the non-redundant rainbow skip graph to support ill-mannered node deletions, or node *failures*, in which a node leaves the network without first notifying its neighbors and providing the necessary information to update the graph structure. More significantly, this scheme will, with high probability, allow us to efficiently restore the proper structure even if all nodes simultaneously fail independently with some constant probability less than one. In particular, we will be capable of reconnecting components of the graph that are disconnected after the failures. We refer to this augmented data structure as the *rainbow skip graph*.

---

[*]For the sake of simplicity, we assume that the nodes in our network are synchronized; we address in the full version the concurrency issues that relaxing this assumption requires.

## 2.1 The Structure of Non-Redundant Rainbow Skip Graphs

A non-redundant rainbow skip graph on $n$ nodes consists of a skip graph on $\Theta(n/\log n)$ *supernodes*, where a supernode consists of $\Theta(\log n)$ nodes that are maintained in a doubly-linked list that we will refer to as the *core list* of the supernode. The nodes are partitioned into the supernodes according to their keys so that each supernode represents a contiguous subsequence of the ordered sequence of all keys. The smallest key of a supernode $S$ will be referred to as *the key of $S$*, and we use these keys to define the skip graph on the supernodes. For each supernode $S$, we associate a different member of $S$ with each level $i$ of the skip graph, and call this member *the level $i$ representative of $S$*, which we denote as $S_i$. The level $i$ list to which $S$ belongs will contain $S_i$. Collectively we refer to these lists of the skip graph as the *level lists*. $S_i$, which can be chosen arbitrarily from among the elements of $S$, will be connected to $S_{i+1}$ and $S_{i-1}$, which we respectively call the parent and child of $S_i$. These vertical connections form another linked list associated with supernode $S$ that we refer to as the *tower list* of $S$. By maintaining the supernodes so that their size is greater than their height in the skip graph, each member of a supernode will belong to at most three lists—the core list, the tower list, and one level list. The issue of supernode size and height will arise frequently, and we let $S.size$ and $S.height$ denote the size and height of $S$, respectively. The implicit connections between the nodes in a core list and their copies in the related tower list form one type of "rainbow" connections, which motivates the name of this structure. (See Figure 3.)

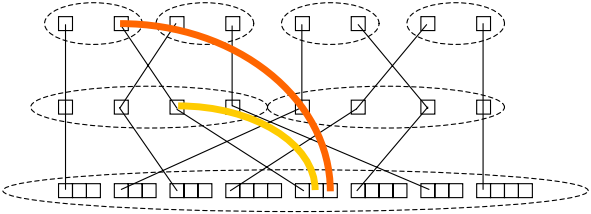

Figure 3: A non-redundant rainbow skip graph. The lists on each level are shown in dashed ovals. The core lists on the bottom level are shown as sets of contiguous squares. The rainbow connections between one core list and its related tower list are also shown.

**Searching in a Non-Redundant Rainbow Skip Graph.** To search for a node with key $k$ from node $x$, we find the top-level representative of the supernode of $x$ and then perform a standard skip graph search for the predecessor of $k$ in the set of supernode keys. Once the predecessor of $k$ is found, we linearly scan through the corresponding supernode until a key with value $k$ or more is encountered, and return the address of the corresponding node to node $x$. Each of these steps requires $O(\log n)$ time, given that we properly maintain the size of every supernode to be $O(\log n)$. We subsequently address this maintenance in the discussion of insertion and deletion operations.

**Updating a Non-Redundant Rainbow Skip Graph.** The method of maintaining supernode sizes of $O(\log n)$ is essentially the standard merge/split method such as that used in maintaining

B-trees—we set constants $c_1$ and $c_2$ such that the size of a supernode is always between $c_1 \log n$ and $c_2 \log n$, merging two adjacent supernodes whenever one drops to a size less than $c_1 \log n$, and splitting a supernode into two supernodes whenever it reaches a size of more than $c_2 \log n$. The primary complication with this approach stems from the distributed setting, in which one cannot efficiently maintain the exact value of $n$ at every node—to do so would require a message to every supernode upon every insertion. The common solution is to estimate the value of $\log n$ locally via some random process.

With some slight modifications to a proof of a theorem from [21] we can arrive at the following.

**Theorem 1:** *In a skip graph on $n$ nodes, the height of every node is $\Theta(\log n)$ with high probability.*

**Proof:** Consider first the probability that the height of some particular node $x$ is $k_1 \log n$ or less. This happens only if all other nodes differ from $m(x)|k_1 \log n$ in at least one membership vector bit. The probability of this is $[1 - (\frac{1}{2})^{k_1 \log n}]^{n-1} = (1 - \frac{1}{n^{k_1}})^{n-1} \leq 2(1 - \frac{1}{n^{k_1}})^n < e^{\frac{-n}{n^{k_1}}} = e^{-n^{1-k_1}}$. Thus the probability that there exists some node with height $k_1 \log n$ or less is at most $ne^{-n^{1-k_1}}$. Consider now the probability that the height of some particular node $x$ is $k_2 \log n$ or more. This happens if one or more of the other nodes agree with all bits of $m(x)|k_2 \log n$. This probability is no more than $(n-1)(\frac{1}{2})^{k_2 \log n} < \frac{1}{n^{k_2-1}}$. It follows that the probability that there exists some node with height of $k_2 \log n$ or more is at most $\frac{1}{n^{k_2-2}}$. Suitably chosen values of $k_1$ and $k_2$ will yield a suitably-high probability. ∎

Ideally we would simply use *S.height* as the estimate for $\log n$. However, the height of a node can potentially change dramatically when its neighbor at the highest level is deleted, or when a new neighbor is inserted. This creates the potential for a cascading series of supernode merges and splits due solely to changes in this local probabilistic estimate of $\log n$, which complicates the otherwise-straightforward amortization argument. For simplicity, we deal with this complication by maintaining an estimate $\log n'$ that is common to every node, in the following manner: whenever some supernode has a height that is outside of the range $[\frac{1}{3} \log n', 6 \log n']$, we recompute the current number of nodes in the structure, $n''$, (requiring $\Theta(n)$ messages), set $n' = n''$, and rebuild the entire structure. With high probability, Theorem 1 guarantees that we rebuild only after the size of the structure has increased from $n'$ to $(n')^2$ or has decreased to nearly $(n')^{1/3}$. In either case, this implies that, with high probability, $\Omega(n'')$ operations have been performed, which will suffice to yield an amortized cost of $O(\log n)$.

We can now describe the insertion procedure. To insert a node $x$ with key $k$ we first search for the predecessor of $k$ and insert $x$ into the corresponding supernode $S$. If *S.size* exceeds $9 \log n'$, then we split $S$ into two equal-sized supernodes. The supernode containing the larger keys of $S$ must then be inserted into the skip graph; it can be inserted by the standard insertion procedure of a skip graph, except that at each level, a different representative is inserted into the corresponding list. We omit the details of this operation as it is a relatively straightforward adaptation of the skip graph method.

Similarly, if, upon deleting a node, *S.size* falls below $3 \log n'$, then we merge $S$ with one of it's neighbors, or simply transfer a fraction of the neighbor's nodes to $S$ if the total number of nodes between them exceeds $9 \log n$. If $S$ is merged with its neighbor, then the old supernode is deleted from the skip graph.

**Load Balance in a Non-Redundant Rainbow Skip Graph.** The following theorem bounds the congestion of a non-redundant rainbow skip graph.

**Theorem 2:** *The congestion of an $n$ node non-redundant rainbow skip graph is $O(\log n/n)$.*

**Proof:** There are three phases to the search: first, traversing the tower list to the top-level representative; second, traversing the level-lists to find the supernode with key nearest the destination; third, to traverse the core list of this supernode until finding the actual goal. Only an $O(\frac{\log n}{n})$-fraction of queries will pass through a particular node $u$ with key $k(u)$ in phases one and three because only that fraction of keys belong to the same supernode as $x$. To analyze the fraction of queries that pass through $u$ in the second phase, we consider first the probability that a search with a particular start and destination pair (s,t) passes through node $u$, and then bound the average over all choices of (s,t). Let $i$ be the level at which $u$ is a representative in the skip graph. Note that a query from $s$ to $t$ will pass through node $u$ only if the following two conditions hold:

1. $m(s)|i = m(u)|i$

2. $\nexists v$ s.t. $k(u) < k(v) \leq k(t)$ and $m(v)|i+1 = m(s)|i+1$;

that is, there is no node between $u$ and $t$ that is present in level $i+1$ of the start's skip list.

These events are independent and thus the probability that they hold is exactly $\frac{1}{2^i}(1 - \frac{1}{2^{i+1}})^d$, where $d$ denotes the number of supernodes whose keys are between $k(u)$ and $k(t)$. Note that the particular choice of $s$ plays no role in this probability. We thus average only over the choices of $t$. Noting further that for each distance $d$ there are only $O(\log n)$ nodes whose supernodes fall at a distance of exactly $d$, we change the summation to be over $d$, yielding the following bound on the congestion of $u$:

$$\begin{aligned} congestion(u) &\leq \frac{c \log n}{n} \frac{1}{2^i} \sum_{d=0}^{\infty} (1 - \frac{1}{2^{i+1}})^d \\ &\leq \frac{c \log n}{n} \frac{1}{2^i} 2^{i+1} \quad \text{(geometric series)} \\ &= \frac{2c \log n}{n}. \end{aligned}$$

∎

In the section that follows we introduce a novel substructure that we term *hydra components*. Members of a hydra component will remain connected with high probability even if a constant fraction of the members simultaneously fail. By employing erasure-resilient codes, two hydra components can be "linked" so that the non-failing members of each component can remain connected to each other. We will use this to partition the lists of the non-redundant skip graph into hydra components and link them via "rainbow connections" in such a way that all non-failing representatives of a given level will remain connected to each other, and such that the representatives of level $i$ will remain connected to representatives of level $i + 1$. Collectively this will ensure total connectivity and provide a simple and efficient means of locally correcting the link structure of the rainbow skip graph.

## 3  Hydra Components

We now describe hydra components—collections of nodes organized in such a way that if each member fails independently with constant probability $p$, then with high probability the nodes that remain can collectively compute the critical network-structure information of all the nodes in that component, including those which have failed. This "critical information" should consist of whatever information is necessary in order to remove local failed nodes from the overlay network and recompute correct links for the nodes that remain. Although in principal these failure-resilient blocks can be designed to handle any constant failure probability less than 1, for simplicity we define them to handle a failure probability of $1/2$.

An $(n, c, l, r)$-erasure-resilient code consists of an encoding algorithm and a decoding algorithm. The encoding algorithm takes a message of $n$ bits and converts it into a sequence of $l$-bit packets whose total size is $cn$ bits. The decoding algorithm is able to recover the original message from any set of packets whose total length is $rn$. Alon and Luby [1] provide a deterministic $(n, c, l, r)$-erasure-resilient code with linear-time encoding and decoding algorithms with $l = O(1)$. Although these codes are not generally the most practical, they give the most desirable theoretical results.

To build the hydra-components, we make use of a $2d$-regular graph structure consisting of the union of $d$ Hamiltonian cycles. The set of all such graphs on $n$ vertices is denoted by $H_{n,d}$. A $(\mu, d, \delta)$-hydra-component consists of a sequence of $\mu$ nodes logically connected according to a random element of $H_{\mu,d}$, with each node storing an equal share of a message encoded by a suitably-chosen $(n, c, l, r)$-erasure-resilient code. The parameters of the erasure-resilient code should be chosen in such a way that the entire message can be reconstructed from the collective shares of information stored by any set of $\delta\mu$ nodes, i.e., such that $\frac{r}{c} = \delta$. The message that is encoded will be the critical information of the component. Clearly if the critical information consists of $M$ bits, then by evenly distributing the packets of the encoded message across all the nodes, $O(M/\mu)$ space is used per node, given that $\delta$ is a constant. In addition to this space, $O(\mu)$ space will be needed to store structures associated with the erasure-resilient code. However, in our applications $\mu$ will be no larger than the space needed for $O(1)$ pointers, i.e., no more than $O(\log n)$.

To achieve a high-probability bound on recovery, we rely upon the fact that random elements of $H_{\mu,d}$ are likely to be good expanders. In particular we make use of a theorem from [8], which is an adaptation of a theorem from [7, 6]. We state the theorem below, without proof.

**Theorem 3:** [8] Let $V$ be a set of $\mu$ vertices, and let $0 < \gamma, \lambda < 1$. Let $G$ be a member of $H_{\mu,d}$ defined by the union of $d$ independent randomly-chosen Hamiltonian cycles on $V$. Then, for all subsets $W$ of $V$ with $\lambda\mu$ vertices, $G$ induces at least one connected component on $W$ of size greater than $\gamma\lambda\mu$ with probability at least

$$1 - e^{\mu[(1+\lambda)\ln 2 + d(\alpha \ln \alpha + \beta \ln \beta - (1-\lambda)\ln(1-\lambda))] + O(1)},$$

where $\alpha = 1 - \frac{1-\gamma}{2}\lambda$ and $\beta = 1 - \frac{1+\gamma}{2}\lambda$.

With suitably chosen $\mu$, $\lambda$, and $\gamma$, Theorem 3 will guarantee that with high probability at least $\gamma\lambda\mu$ nodes are connected, conditioned on the event that $\lambda\mu$ nodes of the component have not failed. By applying Chernoff bounds, it can be shown that this event occurs with high probability for suitably chosen $\mu$ and $\lambda$. These facts directly yield the following lemma and theorem.

**Lemma 1:** If each node of a component of $\beta \log n$ nodes fails independently with probability $\frac{1}{2}$, then the number of non-failing nodes is no less than $\lambda \mu$ with probability at least

$$1 - \left(\frac{e^{\delta}}{(1+\delta)^{(1+\delta)}}\right)^{\frac{\beta \log n}{2}}$$

where $\delta = 1 - 2\lambda$. **Proof:** Let $X$ be the sum of failed nodes. With $\mu = \beta \log n$ and the failure probability $= \frac{1}{2}$, $E(X) = \frac{\beta \log n}{2}$. Since each node fails independently, a Chernoff bound can be applied to the probability that more than $(1-\lambda)\mu$ nodes would fail:

$$\begin{aligned} Pr[X > (1-\lambda)\mu] &= Pr[X > (1+\delta)\frac{\beta \log n}{2}] \\ &< \left(\frac{e^{\delta}}{(1+\delta)^{(1+\delta)}}\right)^{\frac{\beta \log n}{2}}. \end{aligned}$$

∎

**Theorem 4:** For any constant $k$, there exist constants $d$, $\beta$, and $\delta$ such that with probability $1 - O(\frac{1}{n^k})$ the critical information of a $(\beta \log n, d, \delta)$ hydra component can be recovered in $O(\log n)$ time when each node of the component is failed with probability $\frac{1}{2}$.

**Proof:** Setting $\beta$ to 40 and $\lambda$ to $\frac{3}{10}$ provides a lower bound of $1 - O(\frac{1}{n^2})$ on the probability that at least $\frac{3}{10}$ of the nodes do not fail. Theorem 3 can now be applied, with $\mu = \beta \log n$, $\gamma = \frac{1}{2}$, and $d = 47$, to guarantee that there is a connected component amongst the $\frac{3}{10}$-fraction of unfailed nodes of size $\frac{3\mu}{20}$ with probability $1 - O(\frac{1}{n^2})$. Thus the probability that the conditions of Theorem 3 and Lemma 1 both hold is $1 - O(\frac{1}{n^2})$ for the given values of the parameters. This can be extended to any other value of $k$, for example by changing $\beta$, which appears in the exponent of both probability terms, by a factor of $\frac{k}{2}$.

By employing an intelligent flooding mechanism, the packets held by the $\frac{3\beta \log n}{20}$ connected nodes can be collected together in $O(\log n)$ messages. The erasure-resilient code can then be used to reconstruct the critical information by choosing parameters of the code so that $\delta = \frac{r}{c} = \frac{3}{20}$. ∎

For the remainder of the paper, when we use the term "hydra component" we will implicitly mean the $(\Theta(\log n), \Theta(1), \Theta(1))$ hydra components described in the proof of Theorem 4 unless otherwise stated.

## 4 Rainbow Skip Graphs

Having defined a hydra component, what remains is to describe how the nodes of a non-redundant skip graph are partitioned into hydra components to yield the complete rainbow skip graph.

Let $9\beta \log(n')$ be the minimum size of each hydra component, which will be at least $\beta \log n$ with high probability. The maximum size of a hydra component will be maintained as $27\beta \log(n')$. The elements of the level lists are partitioned into hydra components with respect to their order in the lists — elements of a contiguous sublist will be placed together in a hydra component. We call such hydras the level-list hydras. Naturally, sometimes a list, or the remaining end of a list, will be too small to fill an entire hydra component. In such cases, the partition will span into the beginning of the next list of the same level. That is, if a hydra component containing the end of the $jth$ list of level $i$ is smaller than $9\beta \log(n')$, then the hydra component will also contain the beginning of the $(j+1)$-th list of level $i$, or the 0th list of level $i-1$ if $j = 2^i - 1$. The core lists and tower lists will be partitioned in a similar manner, but we group the core list and tower list of each supernode together as a pair, since they are inherently tied together through the supernodes, one being a subset of the other. We call these hydras the supernode hydras. Again when the hydra is too small, supernodes adjacent with respect to their keys are grouped together.

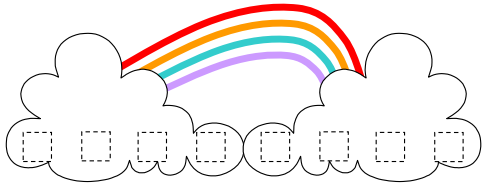

Figure 4: Rainbow connections between hydra components.

The primary critical information that will be associated with every hydra component is an ordered list of the addresses of every node in the component. This information will ensure that the unfailed nodes of a component can restore connectivity to each other locally. In the case of the level-list hydras, the critical information will also include the addresses of the parents and children of every element. Additionally, every pair of adjacent level-list hydras $L$ and $R$ will be *linked* together by "rainbow connections", which amount to storing the addresses of all elements of $R$ in the critical information of $L$ and vice-versa. (See Figure 4.) In total the critical information consists of $O(\log n)$ pointers, which when distributed evenly as encoded packets to the $\Theta(\log n)$ members requires space corresponding to $O(1)$ pointers.

Hydra components are maintained in the same way as supernodes, with the same merge/split mechanism. By design, the cost of a split or merge amortizes to a constant amount of overhead. Whenever a node is added or removed from a hydra component, we must recompute and redistribute the encoded critical information of that hydra and all (of the constant number of) hydras to which it is linked, requiring time proportional to the size of the hydra. Noting that every node in the rainbow skip graph belongs to a constant number of hydra components, we arrive at the following.

**Theorem 5:** *The amortized message cost of insertion and well-behaved deletion, $U(n)$, in a hydra-augmented rainbow skip graph is $O(\log n)$ with high probability.*

**Proof:** The addition of hydras will not affect the global rebuilding costs, so we look only at the cost associated with normal insertions into a supernode $S$. When $S$ does not split or merge during insertion, only the supernode hydra associated with $S$ changes, which requires $O(\log n)$ time. Deletion may also entail a change at a single level-list hydra when a representative of $S$ is being deleted, again incurring only $O(\log n)$ cost. When $S$ does split, we must insert a supernode into the skip graph and potentially must change the representatives of $S$ so that they all come from the lower half of the range of keys in $S$. This will in total change $O(\log n)$ hydras, requiring $O(\log^2 n)$ time. Similar reasoning applies for merging. However, a supernode splits or merges only after $\Omega(\log n)$ insertions and deletions, and thus the amortized time is $O(\log n)$. ∎

We now describe the procedure to restore the structure after some number of nodes have failed. For simplicity we assume that no additional failures occur during the repair process; otherwise, it would be necessary to extend the model to reflect the rate at which nodes are failing with respect to time.

We initially assume that no supernode drops below the minimum size constraints. The goal will be to replace each failed representative with a new node from the same supernode. In parallel, each hydra $H$ in which at least one node has failed recovers its critical information. The core-list information of each supernode $S$ is first used to determine new representatives at each level $i$ with a failed representative. This new $S_i$ is then linked to $S_{i-1}$ and $S_{i+1}$.

With the parent/child information now corrected, what remains is to repair the sibling pointers of each level list. As a basis, we repair level 0 by using the rainbow connections to identify some unfailed member of each supernode that neighbors a failed representative $S_0$. These nodes are sent replacement messages that indicate the new $S_0$, which is then connected to the appropriate neighbors.

The other lists are then repaired sequentially by order of increasing level. Suppose that the lists of level $i$ are being repaired, and hence that the lists at levels 0 through $i-1$ have already been repaired. Let $S$ be some supernode which contains a failed representative at level $i$. To restore the proper structure, we wish to pass an insertion message to the neighbors of $S$ at level $i$ that contains the address of the new $S_i$. We use $S_{i-1}$ to enter the list at level $i-1$ and pass replacement messages to its left and right neighbors, which forward the messages until nodes belonging to the same level $i$ list are reached. With high probability the distance to these nodes is $O(\log n)$.

Once all levels are repaired, the hydra codes are recomputed, completing the repair process. Careful accounting of all actions described above yields the following theorem.

**Theorem 6:** *The failure recovery procedure restores the correct structure of the rainbow skip graph with high probability using $O(\log^2 n)$ rounds of message passing and $O(min(n, k \log n))$ messages, where $k$ is the number of nodes that failed.*

## 5 Strong Rainbow Skip Graphs

The rainbow skip graph achieves $O(\log n)$ query time w.h.p. Although the random construction of the hydra components is critical for the rainbow skip graph to be failure-resilient, we are still able to de-randomize other parts of the structure to get an efficient worst-case search time. The resulting data structure, which we call the *strong rainbow skip graph*, will function as a deterministic peer of the family tree [21] (which is non-trivial to de-randomize), and will additionally provide powerful failure-resilience. The idea is to integrate the randomly constructed hydra components into a deterministic non-redundant rainbow skip graph in the same way it is integrated into a randomized non-redundant rainbow skip graph. This can be done once we have a deterministic skip graph at hand, noting that the method of partitioning and constructing hydra components is independent of the underlying skip graph. Doing this will, in addition to guaranteeing a worst case search time, yield a tighter size for the supernodes and more freedom to rebuild than that of the ordinary rainbow skip graph. We discuss the size of the supernodes and that of the underlying deterministic skip graph in the next, then provide a novel deterministic skip graph in Section 5.1, and present how to update this structure in Section 5.2 and 5.3.

**Size of Supernodes.** Let $n$ be the number of keys when the current skip graph was built and $l = \log n$. We maintain a supernode size between $[l, 3l-2]$. If a deletion turns a supernode $S$ into $l-1$ size, then $S$ either borrows a key from the next block $S'$ if there are at least $l+1$ keys in $S'$, or merges with $S'$ into a new supernode of size $2l-1$. Similarly, if an insertion turns $S$ into size $3l - 1$, $S$ either gives the additional key to $S'$ if $S'$ contains at most $3l - 3$ keys, or merges with $S'$ and then splits into three supernodes of size $2l - 1$. Thus any newly generated supernode is of size $2l - 1$, so it will tolerate at least $l$ insertions or deletions before the next merge or split.

**Size of The Skip Graph.** A newly rebuilt skip graph has $n$ keys and $n/\log n$ supernodes. We can rebuild it as soon as it grows to $n' = n$ or shrinks to $n' = n/2 \log n$ supernodes. (As in the randomized case, the number of supernodes can be estimated by the height of each tower list so there is no need of any global information.) This provides that before a rebuild there have been at least $(n' - n'/\log n)$ supernode insertions or $n'$ supernode deletions, so that the time for rebuild, which is $O(n' \log n)$, upon being amortized, requests only constant credits from each key insertion and deletion. On the other hand, in order to keep the $O(\log n)$ search time, the current skip graph can accommodate as many as $n'' = n^c$ or as few as $n'' = n^{1/c}$ supernodes. Therefore we can rebuild at any point between $n'$ and $n''$ without affecting either the amortized update time or the worst-case search time.

As summarized in the following theorem, we are left with the task of building a deterministic skip graph as the underlying structure of the strong rainbow skip graph.

**Theorem 7:** *If there is a deterministic skip graph with worst-case query time $Q$, worst-case update time $U$, and congestion $C$, then there is also a strong rainbow skip graph with worst-case query time $Q$, amortized update time $U$, congestion $C$, and resilience to node failures with constant probability.*

## 5.1 A Deterministic Skip Graph with Near Optimal Congestion

Harvey and Munro [11] gave a deterministic SkipNet with $O(\log n)$ search time and $O(\log^2 n)$ update time, which could be used in the construction in Theorem 7. However each node in [11] has three parents and two to five children so that the congestion at a top level node could be as bad as $(1/3)^{\log_5 n} = n^{-\log_5 3} = n^{-0.68}$. Here we provide another deterministic skip graph with the same search and update time but only $(1/2)^{\log_{2(k+1)/k} n} = n^{-1/(1+\log((k+1)/k))}$ congestion, with $k$ being a free parameter, where each node has two parents and only $(2 + 1/k)$ children in average. Compared with the optimal congestion $O(\log n/n)$ that is only achieved by randomized structures, our deterministic skip graph has $O(n^\epsilon/n)$ congestion, which is the first deterministic p2p structure to provide such near optimal load balance.

**Macro Structure.** Like in a randomized skip graph or SkipNet, the macro structure of our new deterministic skip graph is still an upside-down tree consisting of sublists, where there are $2^i$ sublists at the $i$-th level. Each sublist has a father (0) list and a mother (1) list in the level above it. Each key $x$ still has a membership string whose first $i$ bits determine the sublist at level $i$ to which the copy of $x$ at this level belongs. However, some bits of the string might be *blurred*, by which we mean that $x$ is present but not counted at these levels when we balance the skip graph. Furthermore, some bits might be *undetermined*, meaning that $x$ is not yet inserted into these levels and they hence have no value. Naturally, if bit $i$ is undetermined, then any bit $j > i$ is also undetermined.

The balance property of this skip graph is then that, in any sublist at any level, consecutive non-blurred nodes promote to the father list and the mother list alternatingly, and the blurred nodes are spread more than $k$ steps apart. We may think of these normal nodes as promoting to the next level in pairs, where in each pair of sibling nodes one goes to the father and the other goes to the mother. And we may call this pair of siblings the *dual* of each other. As defined, a blurred node may promote to either the father list or the mother list, or neither of them (when the next digit is undetermined). That is, the deterministic skip graph deviates from a perfect structure, where consecutive nodes in each list promote to father list and the mother list alternatingly, by tolerating some sparsely spread blurred nodes.

**Theorem 8:** *(Height is near perfect.)* The height of any tower in the skip graph is between $\frac{\log n}{1+\log((k+1)/k)}$ and $\frac{\log n}{1+\log((k+1)/(k+2))}$.

**Proof:** If there are $m$ normal nodes and 0 to $m/k$ (the max. possible) blurred nodes at level $i$, then there are at most $(m/2 + m/k)$ and at least $m/2$ nodes in either of its parent lists at level $i+1$. So the shrinking ratio from the size of a child list to that of a parent list is between $(1+1/k)/(1/2+1/k) = 2(k+1)/(k+2)$ and $(1+1/k)/(1/2) = 2(k+1)/k$, so that the height of any skip list inside this skip graph is between $\log_{2(k+1)/k} n = \frac{\log n}{1+\log((k+1)/k)}$ and $\log_{2(k+1)/(k+2)} n = \frac{\log n}{1+\log((k+1)/(k+2))}$. ∎

**Theorem 9:** *(Congestion is near optimal.)* The maximum congestion at any tower, i.e., the congestion of the skip graph, is at most $O(\log n / n^{1-\epsilon'}) = O(n^\epsilon / n)$ with $\epsilon$ only depending on $k$.

**Proof:** We calculate the congestion at each node and then sum it for each tower. Let S and T be the start and destination of randomly chosen search. In a perfect skip graph with two parents and two children for each node, the congestion at a node $x_i$ at level $i$ is the probability of $S$ being inside the left or right subtree of $x_i$ (which is $2^i/n$) times the probability of $T$ being outside that subtree (relaxed to 1 and ignored) and then times the probability of choosing the right search tree involving $x_i$ (which is $1/2^i$), so it is $1/n$. In our skip graph, the probability of $S$ being inside a subtree of $x_i$ is at most $(2(k+1)/k)^i/n$, so the congestion is $O((1/2)^{\log_{2(k+1)/k} n}) = O(n^{-1/(1+\log((k+1)/k))})$ by repeating the above calculation. ∎

Next we show how to update this skip graph to maintain the balance property. We first provide update operations with amortized $O(\log n)^2$ time, then improve the running time to amortized $O(\log n)$ by grouping nodes into supernodes in a different way.

## 5.2 Update Operations in The Deterministic Skip Graph

As described above, the deterministic skip graph is an approximation to a perfect skip graph, with a balance constraint the any two blurred nodes in a same list are separated at least $k$ steps away. When this is violated during updates and blurred nodes become too dense, we follow the procedure below to restore the balance property. A straightforward case is that, if there are two adjacent neighbors in a list that are both blurred, and they promote to the father and mother list alternatively and consistent to the alternation of other nodes in the same list, then we can simply unmark them as blurred nodes (unblur) and consider them two normal nodes. Then, if two blurred nodes are within $k$ steps, we can do local surgeries to the structure to yield similar result as in the straightforward case. Details are as follows.

**Swap.** To swap two neighbor nodes at level $i$ means to exchange their membership strings from the digit $i+1$ above (as long as the digits are determined, including the blurred digits). This changes neither the structure of the skip graph nor the valid search property, since the two being neighbors at level $i$ means that one can replace the other at any level above $i$.

**Unblur or Promote in Pairs.** If, in any sublist, there are two blurred nodes within $k$ steps away, and they promote to different parents, then we can unblur them both and fix the balance property, if it is violated, by $O(k)$ swaps. This works regardless of whether there is an odd or even number of normal nodes in between of the two blurred nodes. For example, if we use $f$ and $m$ to indicate that a node promotes to father list or mother list, and $f(b)$ (or $m(b)$) means it is blurred and promotes to the father (or mother) list, then a sublist $[x_1 = f, x_2 = f(b), x_3 = m, x_4 = f, x_5 = m, x_6 = m(b), x_7 = f]$ will be fixed by swapping $(x_2, x_3)$, $(x_4, x_5)$ after unblurring $x_2$ and $x_6$, and a sublist $[x_1 = f, x_2 = f(b), x_3 = m, x_4 = f, x_5 = m, x_6 = f, x_7 = m(b), x_8 = m]$ will be fixed by swapping $(x_2, x_3)$, $(x_4, x_5)$ and $(x_6, x_7)$ after unblurring $x_2$ and $x_7$. A blurred node with the next digit undetermined can be promoted together with another blurred node regardless of which parent the latter promotes to. If there are two blurred nodes going to the same parent, then we cannot unblur them together no matter how close they are. However, it will turn out that a node is blurred only during the deletion, and that we can control the deletion so that if a node is going to be blurred and there exists another blurred node nearby, we can blur either the node or its dual so that the newly blurred node can always be coupled with the existing blurred one. (See the *deletion* below.)

The unblurring of a blurred node with its next digit determined doesn't cause any change to the upper level. The promotion of a blurred node with the next digit undetermined, which unblurs this node and then inserts a blurred node (determines a bit) at the upper level, will propagate if the newly inserted node is within $k$ steps to another blurred node in the same sublist.

**Insertion.** To insert a key we first insert into the bottom list a blurred node and assign to it an empty membership string with all digits undetermined. We then propagate the promotions until the balance property is restored.

**Deletion.** We delete all copies of the key $x$ from the bottom level up to the top level. At each level $i$, in the first case, if there is no blurred node within $k$ steps from $x_i$ or its dual, then delete $x_i$ and blur its dual. In the second case, if there is a blurred node $y_i$ within $k$ steps that promotes to the same parent as $x_i$, then delete $x_i$ and unblur $y_i$, and accordingly fix the segment between $y_i$ and the dual of $x_i$ by $O(k)$ swaps. In the last case, if the blurred node $y_i$ within $k$ steps promotes to a different parent than $x_i$, then swap $x_i$ with its dual to make it the second case.

**Theorem 10:** *The amortized message cost of each insertion and deletion in the deterministic skip graph is $O(\log^2 n)$.*

**Proof:** Each swap takes $O(\log n)$ messages since the two strings have $O(\log n)$ digits to exchange and exchanging a digit results in $O(1)$ pointer changes in the data structure. To each insertion we assign $O(\log^2 n)$ credits to the inserted blurred node at the bottom level, $O(\log n)$ credits per undetermined digit in its membership string. For a deletion, at each level it takes $O(k)$ swaps and blurs one node, to which we also assign $O(\log n)$ credits. Thus an insertion/deletion takes $O(\log^2 n)$ messages and each existing blurred node (bit) or undetermined digit carries $O(\log n)$ credits to afford for the future propagation of unblurrings/promotions. ∎

### 5.3 Getting $O(\log n)$ Update Time in The Strong Rainbow Skip Graph

A direct grouping of keys into supernodes and integration of hydra components into the above deterministic skip graph results in a rainbow skip graph with amortized $O(\log^2 n)$ update time, as mentioned in Theorem 7. To improve the update cost to $O(\log n)$, we propose a two-level grouping instead of the (one-level) grouping described in Section 2. Instead of grouping the keys into supernodes of size $O(\log n)$, we now group them into supernodes of size $O((\log^2 n))$, and further partition each supernode into a list of $O(\log n)$ sublists each of size $O(\log n)$, where the keys in the first sublist are all smaller than the keys in the second list, etc. This way the search/query time remains $O(\log n)$, because after a supernode containing the query key is located, to search for the key inside the supernode we only perform two levels of linear traverse of lists, each of size $O(\log n)$. With splitting and merging the sublists inside a supernode in the same way a supernode is split or merged, we should be performing a supernode insertion or deletion only once every $\Theta(\log^2 n)$ insertions or deletions of keys, resulting in an update cost amortized to $O(\log n)$. To deploy in a distributed environment, we associate each sublist in a supernode to the $i$-th level representative of this supernode in the skip graph, and use any of the keys in the sublist to be the representative. By maintaining $O(\log n)$-sized hydra components, we can follow the same argument to see that the amortized update time after integrating hydra components is still $O(\log n)$. This two-level grouping scheme will bring one more factor of $\log n$ to the congestion because the number of keys associated to a supernode has increased by a factor of $\log n$. Theoretically, however, this factor of $\log n$ is covered by the $n^\epsilon$ in Theorem 9. Moreover, since now we have a sublist of $O(\log n)$ candidate representatives for each node in the skip graph and only one of them is used, in practise we are able to cancel this increase of congestion by rotating the representative among the $O(\log n)$ candidates after a search passes this node. Thus we achieve (deterministic) amortized $O(\log n)$ update cost, adding no affect to other features of the rainbow skip graph.

**Theorem 11:** *The amortized message cost of insertion and well-behaved deletion, $U(n)$, in a strong rainbow skip graph is $O(\log n)$.*

## Acknowledgments

This research is supported in part by the NSF under grants IIS-0713046, CCR-0830403, and the Office of Naval Research under MURI Award number N00014-08-1-1015. This paper appeared in preliminary form in the ACM-SIAM Symposium on Discrete Algorithms as [10].